\documentclass[12pt,aps,prd,preprint,showkeys]{revtex4}
\usepackage[utf8]{inputenc}
\usepackage{amsmath}
\usepackage{amsfonts}
\usepackage{amssymb}
\usepackage{graphicx}
\usepackage{csquotes}
\RequirePackage[T1]{fontenc}

\begin{document}

\title{Magnetic field amplification and decay in cosmic string wakes}

\author{Deepanshu Bisht${^1}$}        
\author{Dilip Kumar${^2}$}       
\author{Soumen Nayak${^2}$}       
\author{Soma Sanyal${^2}$}


\affiliation{$^1$Indian Institute of Technology, Kanpur, U.P, India}

\affiliation{$^2$School of Physics, University of Hyderabad,  Telangana, India.}


\begin{abstract}
We do a detailed study on the evolution of a magnetic field in a magnetized plasma within the spacetime of a moving cosmic string using analytical and numerical methods. The conical spacetime around the cosmic string causes the frozen-in magnetic field to deform due to the fluid flow. We find that the overdensity in the wake region amplifies the magnetic field. This amplification depends on the direction and the lengthscale of the magnetic perturbations. Alfven’s theorem of flux conservation explains this result. However, our study also shows that the magnetic field can decay depending on the perturbation lengthscale, due to the breakdown of Alfven’s theorem at a certain lengthscale. We have identified this lengthscale as the gyroradius of the charged particles in the plasma. Our findings are significant for understanding magnetic reconnection in cosmic string wakes.
\end{abstract}
\maketitle

\section{Introduction}
Cosmic strings are topological defects that are formed from the symmetry-breaking phase transitions in the early universe. An interesting part of these defects is the nature of the spacetime around the string. It is known that the cosmic string metric is globally conical and has a deficit angle. Due to the nature of the metric, the geodesics of particles close to the cosmic string move inwards towards the cosmic string. This gives rise to a wake structure behind a moving cosmic string. This wake structure is well studied for an un-magnetized plasma \cite{vishniac,sornborger,stebbins}. It has been found that these wakes leave several signatures during the evolution of the universe. They can give  rise to discontinuities in the Cosmic Microwave Background Radiation (CMBR) \cite{danos}; they may give rise to primordial magnetic fields \cite{vachaspati1,vachaspati2,sovan} and they may also lead to density inhomogeneities which will affect the nucleosynthesis calculations \cite{layek1,layek,sovan1}.  Thus, understanding the structure of these wakes is important for identifying the signatures of cosmic strings in the current universe. Currently, due to the availability of a large amount of data and 
computational resources, there has been a renewed interest in the experimental signature of these exotic objects \cite{Planck}. Some of these are from density inhomogeneities and some from Gravitational Waves (GW) \cite{sousa}. There are also some signatures from synchrotron radiation and 
Gamma Ray Bursts (GRB) \cite{cyr}. Density inhomogeneities are generated in the wakes of cosmic strings. For the generation of density inhomogeneities it is necessary to study the hydrodynamic wake structure. But for signals like synchrotron radiation and GRBs one must consider the magnetic field in 
the background plasma.  The wake structure that has been studied in the literature is essentially the 
hydrodynamic wake structure. The presence of magnetic fields in the wake was generally attributed only 
to the superconducting cosmic strings \cite{rogozin}. Recently, it has been shown that primordial 
magnetic fields can be generated in the wakes of Abelian Higgs strings \cite{sovan} too. So, it is 
possible that wakes of non-superconducting strings will also have a magnetic field. Particles moving 
through these magnetic fields will radiate synchrotron radiation \cite{soumen2}. Such radiation can be 
detected by current detectors. We need to study the electron distribution, the magnetic field 
perturbation and many other parameters before we are able to obtain a distinct signature that points 
to a cosmic string.  Our motivation in this study is to study in greater detail the flow of the 
magnetized fluid around a cosmic string and understand the important lengthscales associated with this 
flow. This is an important factor for the radiation emitted from magnetic reconnections in the cosmic string wake.

In our previous studies, we have shown that multiple shocks can be generated in the cosmic string wake \cite{soumen}. These shocks can collide and release a large amount of energy as radiation. Moreover, it has also been conjectured that the narrow wake structure of the cosmic strings can lead to magnetic reconnection \cite{dilip}. The shocks in the wake are usually enhanced when there is magnetic reconnection and the radiation that is emitted can be of the order of a GRB \cite{DilipGRB}.
 Lengthscales play an important role in magnetic reconnection. One of the requirements of magnetic reconnection is the breakdown of Alfven's theorem in the magnetized plasma. This can happen at small lengthscales. We aim to identify the lengthscale at which magnetic reconnection can occur in cosmic string wakes. This lengthscale is important as it determines the energy released due to magnetic reconnection.

The fluid flow around the cosmic string is driven by the conical metric of the cosmic string. In the pre-recombination era, the magnetic field lines in the early universe are usually frozen into the plasma. Studies have shown that when the plasma undergoes a spherical compression, the magnetic field frozen in the plasma is also affected \cite{king}. Since the flow around the cosmic string results in the fluid density increasing significantly in the wake region, we do a detailed study of the evolution of frozen-in magnetic fields around a cosmic string.

In the literature, the magnetic field in the wake region is generated due to the presence of density inhomogeneities in the plasma. These density inhomogeneities come from the clustering of different particles around the cosmic string. The clustering of these particles mostly depend on their masses \cite{abhisek}.  A long and narrow wake is formed. Previously we have shown in numerical simulations, that when the velocity of the cosmic string is low, the wake structure is diffuse \cite{soumen}. A well developed wake structure is obtained for higher Mach numbers. For simplicity, an ideal fluid approximation is considered without any dissipative forces. To simulate the wake like structure in a locally flat geometry, the knowledge of the boosted metric of a cosmic string is used to initiate a velocity perturbation in the flow of particles as they pass by the cosmic string. 

Though hydrodynamic wakes and shocks are well studied for a cosmic string, it is only in recent times that magnetohydrodynamic shocks have also been studied in the cosmic string wake. Magnetohydrodynamic shocks are difficult to study as the magnetic field lengthscales play a major role in the generation and evolution of the shock \cite{dahlburg}. Among cosmic string wakes, MHD shocks have been discussed previously in ref. \cite{beresnyak}. In recent times, more detailed simulations of such wakes have been carried out in ref. \cite{soumen}. One of the crucial factor, that was not considered previously in these studies was the lengthscale of the perturbations that affect the magnetic and density fields. Lengthscales of both density inhomogeneities as well as magnetic fields affect structure formation and later evolution of large scale structures in the universe \cite{kim}. For cosmic strings, this is particularly important as density perturbation depend on the small-scale structure of the strings \cite{wu} as well as the nature of the particle motion around them \cite{abhisek}. In this respect, our study clearly shows the effect of magnetic perturbations of different lengthscales on the 
evolution of the magnetic field. We find that there is a lengthscale beyond which the magnetic field is not amplified any more. This is the lengthscale at which Alfven's theorem breaks down, Our 
analytical and numerical studies show that this lengthscale corresponds to the gyro radius of the 
charged particle in the magnetic field.

In section II, we discuss the spacetime around a cosmic string and why a moving cosmic string will always generate a velocity perturbation on the particles moving past it. In section III, we discuss the amplification of the magnetic field due to the deformation of the magnetic flux lines. In Section IV, we show the  decay of the magnetic field in presence of a sheared magnetic field. We also show that in the presence of a sheared magnetic field, the amplification or decay depends on the orientation and the lengthscale of the perturbative field. Section V we summarize the work and discuss why it is important to understand the evolution of magnetic fields in the cosmic string wakes for current observational signatures.

\section{The cosmic string metric and wake formation}

The geodesic equations around a cosmic string metric have been studied in detail in the literature \cite{abhisek,hartmann}. In general, particles moving past a cosmic string suffer a change in their velocities close to the cosmic string. This leads to the wake formation behind moving cosmic strings \cite{sornborger}. The metric being symmetric around the deficit angle of the string, the wake structure is also symmetric around the deficit angle of the string. Here, we would like to briefly recapitulate the basics of the metric of a cosmic string and discuss the velocity perturbation on particles going past a long string.  

The simplest metric of a cosmic string in cylindrical coordinates is given by \cite{vilenkin,deruelle},
\begin{equation}
ds^2 =  dt^2 -  d\rho^2 - (1- 4 G \mu)^2 \rho^2 d\theta^2 - dz^2  
\end{equation}
Here, $\mu$ is the mass per unit length of the cosmic string and comes from the symmetry breaking scale at which the cosmic string was formed. A cosmic string always has a finite width. Let us denote the width of the cosmic string by $r_w$. Beyond the width of the cosmic string the exterior metric of the string can be written as, 
\begin{equation}
ds^2 =  dt^2 -  d\rho^2 - \rho^2 d\theta'^2 - dz^2  
\end{equation} 
where, the $\theta'$ now goes from $0$ to $ 2 \pi (1- 4 G \mu)$. In the Cartesian coordinate system, this leads to the standard flat metric, 
\begin{equation}
ds^2 =  dt^2 -  dx^2 - dy^2 - dz^2      \label{Cartesian metric}
\end{equation}
with the definition of $x = \rho ~Cos~ \theta'$ and $y = \rho ~Sin ~\theta'$. This is just the Minkowski space time with a deficit angle. Thus, this means that the standard MHD equations in Minkowski space time can be used to understand the flow around a cosmic string provided the boundary conditions are set so that the angle $\theta'$ is constrained to be within $0$ to $2 \pi (1- 4 G \mu)$. 

However, there is a difference in the case of the moving cosmic string. This difference lies in the fact that a moving particle is subject to  a velocity perturbation towards the cosmic string. It is given by $\delta v \approx \delta \theta v_s \gamma_s$, where, $v_s$ is the cosmic string velocity and $\gamma_s$ is the Lorentz factor corresponding to the moving cosmic string. This is because of the large mass per unit length of the string which is like a gravitational force and can be written as a metric perturbation obtained from the linearized Einstein's equations for the cosmic string, details of this are provided in ref. \cite{vachaspati2}. A particle at rest with respect to a string is not acted upon by any gravitational force but if the string moves with a velocity $v_s$ nearby matter particles are subject to a boost. The exact expression of the velocity perturbation depends  on the nature of the metric and therefore on the particular kind of cosmic string that we are working with. One can obtain the most general form of the boosted metric in the Cartesian coordinate as \cite{Maarten}, 
\begin{equation}
ds^2 =  dt^2 -  dx^2 - dy^2 - dz^2  - 4G \mu (2 - 4 G \mu) 
\left(\frac{(ydx - xdy)^2}{(x^2 + y^2)} \right)
\end{equation} 
In this case, since $4 G \mu << 1$, the last term can be treated as a metric perturbation on the flat space time. However, as has been discussed in detail in 
ref. \cite{rezolla}, this metric perturbation does have a role to play in changing the geodesics of the particles flowing around the cosmic string. It has been shown that for both long strings as well as wiggly cosmic strings this results in a velocity perturbation proportional to the deficit angle of the cosmic string \cite{vachaspati1,vachaspati2}. 

The metric is written as $g_{\mu \nu} = \eta_{\mu \nu} + h_{\mu \nu}$, where  $ \eta_{\mu \nu}$ is the metric for the flat space time and $ h_{\mu \nu}$ is the perturbation as $4 G\mu << 1$. So, the geodesic equation of motion will only be affected by the metric perturbation in the $x-$ and $y-$direction. The equations of motion with the perturbed metric is a rather complex equation which can be solved numerically. The nature of the geodesic equations in the $x-$ and $y-$directions are, 
\begin{equation}
\frac{dv_x}{dt} = -4 G \mu (2-4 G\mu) (\frac{v_x^2}{c} f_1(x,y) + \frac{v_x v_y}{c} f_2(x,y) + \frac{v_y^2}{c} f_3(x,y))  
\end{equation}
\begin{equation}
\frac{dv_y}{dt} = -4 G \mu (2-4 G\mu) (\frac{v_y^2}{c} f_1(x,y) + \frac{v_x v_y}{c} f_2(x,y) + \frac{v_x^2}{c} f_3(x,y))  
\end{equation}
Here, the functions $f_1$, $f_2$, $f_3$ are functions of the variables $x, y$ and the constants $G, \mu$. They are the functions involving squared and cubic terms and can only be solved numerically but they do indicate that there would be a velocity change on the plasma particles due to the motion of the cosmic string. While studying accretion of particles in cosmic string wakes, a novel method is used to get this velocity perturbation. If a particle has a velocity $\vec{v}$ and it's trajectory has to be computed, it is given a Lorentz transformation to the cosmic string's rest frame and the trajectory is computed in that frame of reference, it is then transformed back to obtain the change in velocity due to the moving string. In ref. \cite{bertschinger}, the velocity change for a relativistic string moving along the $y-$direction has been found. Since the cosmic string in our case is moving along the x-axis, the velocity perturbation due to the moving string in our case would be, 
\begin{equation}
\Delta v_x = - 4 \pi \frac{G \mu \gamma_s}{c^2} (v_s - \frac{v_s^2 v_x}{c^2})    \label{Vx_perturbation}
\end{equation}  
\begin{equation}
\Delta v_y = - 4 \pi \frac{G \mu \gamma_s}{c^2} (v_s - v_x - \frac{v_s v_y^2}{c^2}) \label{Vy_perturbation}
\end{equation}  
Here, $v_s$ is the velocity of the string.

There are different methods to obtain the velocity perturbations due to moving cosmic strings. Since the velocity perturbation depends on the exact metric of the cosmic string, it will be numerically different for different kinds of cosmic string. But the important point that we see in the two calculations is that the velocity perturbation is always proportional to the symmetry breaking scale of the cosmic string i.e.  $\Delta v \propto 4 \pi G \mu$.  So, particles moving past a cosmic string will move in a locally flat space time but will be subjected to a velocity perturbation as they pass by the cosmic string. The metric though flat will have a condition on the angle $\theta$ that the total angle in the space should sum upto $2 \pi (1- 4 G \mu)$ only. Any simulation of cosmic string wakes should satisfy these two conditions.

Wake formation from cosmic string has been studied in hydrodynamics previously \cite{sornborger} and it has been shown that particles with different masses would cluster differently in the cosmic string wakes. In recent times, motion of long cosmic strings have been studied in a magnetized plasma using the magnetohydrodynamic equations in locally flat space time but with a deficit angle \cite{soumen}. Loops of magnetic reconnection has been observed in the simulation. The possibility of having magnetic reconnection in cosmic string wakes have also been shown in a separate work \cite{dilip}. Since magnetic reconnection can lead to various observable signatures, it is important to understand the lengthscales involved in the evolution of the magnetic fields in magnetized cosmic string wakes.     

\subsection{\textbf{Numerical Simulation of flow around a cosmic string}}

For the numerical simulation of fluid flow around a cosmic string, we would like to point out that generally fluid particles in an ideal fluid follow the geodesic of the underlying metric, only in this case the differential equations denoting the flow of the fluid are not solved for individual particles but for a volume of particles as a whole. So it is the fluid density that is used in the equation and not the individual mass and trajectory of each particle. There has been some formulation of the mapping of the geodesics into the hydrodynamic formulation in the literature. The fluid flow around the cosmic string is beyond the core of the cosmic string so in this region $\rho >> r_w$. As the external metric of the cosmic string, determines the flow of the particles in this region, in Cartesian coordinates, it is given by the metric in Eq.(\ref{Cartesian metric}), the MHD equations in these  coordinates would be given by \cite{mkverma}, 
\begin{equation}
\frac{\partial \rho}{\partial t} + \nabla.(\rho \mathbf{v}) = 0 
\end{equation} 
\begin{equation}
\rho \left(\frac{\partial \mathbf{v}}{\partial t} + (\mathbf{v}.\nabla)\mathbf{v} \right) = - \nabla \left(P_{th} + \frac{B^2}{8 \pi} \right) + (\mathbf{B}.\nabla)\mathbf{B}
\end{equation}
\begin{equation}
\frac{\partial \mathbf{B}}{\partial t} + (\mathbf{v}.\nabla)\mathbf{B} = (\mathbf{B}.\nabla)\mathbf{v}
\end{equation}
The important thing to keep in mind is that the angle $\theta$ in the $x-y$ plane is defined only between $0$ to $ 2 \pi (1- 4 G \mu)$.

We use the publicly available OpenMHD code \cite{openmhd} for our simulations. We use the basic code and use open boundary conditions in the direction of the fluid flow. In the basic code, the MHD equations along with the equation of state, relating the hydrodynamic pressure and the density, $P_H = f(\rho)$ are solved using a perturbative method. For the case of magnetohydrodynamics, the total pressure ($P$) consists of the hydrodynamic pressure and the pressure due to the magnetic field \cite{mkverma}.  The fields ($\mathbf{v}$, $\mathbf{B}$, $P$, $\rho$) mentioned previously in the MHD equations  can be split into a background field and a perturbative field as, $\mathbf{v} = \mathbf{v_{0}} + \mathbf{v_{1}}$, $\mathbf{B} = \mathbf{B_{0}} + \mathbf{B_{1}}$, $P = P_{0} + P_{1}$ and $\rho = \rho_{0} + \rho_{1}$. Here, $(\mathbf{v_{0}}$, $\mathbf{B_{0}}$, $P_{0}$, $\rho_{0})$ are the background fields, and $(\mathbf{v_{1}}$, $\mathbf{B_{1}}$, $P_{1}$, $\rho_{1})$ are the perturbative fields. A Galilean transformation is used to shift to the background flow frame, thus, $\mathbf{v_{0}} = 0$, $\mathbf{v_{1}} = \mathbf{v}$ which means that the velocity field $\mathbf{v}$ is the perturbative field. These are then substituted in the MHD equations mentioned previously, and only the first order terms are retained, assuming the perturbations to be small. The linearized equations are given by,
\begin{equation}
\frac{\partial \rho_1}{\partial t} + \nabla.(\rho_0 \mathbf{v}) = 0 
\end{equation} 
\begin{equation}
\rho_0 \left(\frac{\partial \mathbf{v}}{\partial t} - (\mathbf{B_0}.\nabla)\mathbf{B_1} \right) = - \nabla P_1  - \rho_0 \nabla(\mathbf{B_0}.\mathbf{B_1})
\end{equation}
\begin{equation}
\frac{\partial \mathbf{B_1}}{\partial t} - (\mathbf{B_0}.\nabla)\mathbf{v} =  - \mathbf{B_0} \nabla. \mathbf{v}
\end{equation} 

The basic code is available in one and two dimensions. We have used the 2D code and modified it by giving a velocity perturbation in the plane of the wake to the particles as they cross the cosmic string. The fluid flow is taken to be along the $x$-axis. In the $y$-direction, we have used fixed boundary conditions. The cosmic string is aligned with the $z$-axis, it is identified in the flow by the position at which an initial velocity perturbation is given to the plasma particles. Thus, plasma particles crossing a certain position are all given a velocity perturbation in the $x$-and $y$-direction. The velocity perturbation is given by Eq.(\ref{Vx_perturbation}) and Eq.(\ref{Vy_perturbation}). But this velocity perturbation is for each individual particle moving along a geodesic. Assuming that the velocity of the particles are less than the velocity of the cosmic string which moves close to the speed of light, we consider the average velocity perturbation  to be of the order of $4 \pi G \mu v_r$, where $v_r$ is the relative velocity between the particle and the cosmic string. This average velocity is denoted by $v_0$.  To see that the constraint of the angle is maintained in the $x-y$ plane, we take $v_x = -v_0 Cos \theta $ and 
$v_y = - v_0 Sin \theta$. The angle $\theta$ is proportional to the deficit angle. It can neither be zero nor $2 \pi$ as we know from the discussion in the previous section that the total angular deviation is constrained by the deficit angle of the metric. For all other volume elements which are away from the cosmic string, the velocity perturbation is zero. Thus the velocity perturbation is only given when the plasma moves past the cosmic string.  


Initially, for a better understanding of the simulation, a larger deficit angle was taken so that the results are clearly visible. A larger deficit angle will result in a bigger dimension of the wake. However, we prefer to keep the order of the velocity perturbations that we give to the particles of the same order as the velocity perturbation due to the cosmic string metric.  As mentioned before, the deficit angle for an actual cosmic string is very small, in dimensionless parameters the perturbation would be of the order of the deficit angle with a value of $10^{-5}$ \cite{gangui}. If we give a perturbation of the same order to the velocity along both the dimensions of the wake, the lengthscale of the wake region will be of the order of $10^{-5} Mpc$ or smaller.

\subsection{\textbf{Magnetic fields in rotational flows}}
We now consider the role of the magnetic field in the simulation. The magnetic field $B$ has two components (in $x$-and $y$-direction). For a uniform magnetic field, we give fixed values to both components. Charged particles in magnetic field move in rotational motion and generate vorticity.  Generally, the generation of vorticity in magnetohydrodynamics leads to turbulence in the plasma. However, for a turbulence to develop, we need to introduce a constant source for generating vortices at different length scales. As mentioned before, we have a constant background  magnetic field in the simulation so the vorticity is generated at a fixed lengthscale. For turbulence, the plasma has to be non-ideal and the net vorticity need not be conserved. In our simulation, we do not introduce any other potential so the lengthscales of the generated vortices are fixed by the scale of the cosmic string wake and the background magnetic field. Hence no turbulence occurs in our system.

We plot the magnetic field lines in the simulation. The pressure density relation determines the epoch we are studying. Since this is the pre-recombination era, the magnetic field lines are deformed due to the flow of the fluid. We calculate the total magnetic field in the wake region as well as the peak value of the field. As long as the magnetic field remains constant, we do not expect any change in the magnetohydrodynamic flow. The flow remains steady with minor changes due to the velocity perturbation. We proceed to give a perturbation to the magnetic field. Since we would like the field to be amplified, we give a sheared perturbation to the magnetic field as it is previously known that sheared magnetic fields may lead to amplification of the magnetic field \cite{king}. The field is perturbed first along the $x$- direction and then in the $y$-direction, the perturbations are denoted by $B_x$ and $B_y$, respectively. The results of these perturbations are discussed in the next section.

\section{Sheared magnetic fields and field amplification} 
For a frozen in magnetic field, the frozen in condition ensures that the magnitude of the magnetic  field will be related to the density of the plasma. It has been shown that an isotropic collapsing plasma can lead to the amplification of a magnetic field \cite{king}. The argument was that the flux in the frozen in plasma is conserved, so in the collapsing region, as the cross-sectional area decreases, the magnitude of the magnetic field increases. Now, the formation of a wake behind a cosmic string leads to a substantial increase in the density of plasma behind the cosmic string. In this case too, there is the possibility that the magnitude of the magnetic field will be enhanced.

\begin{figure}[h]
	\centering
	\includegraphics[width=\linewidth]{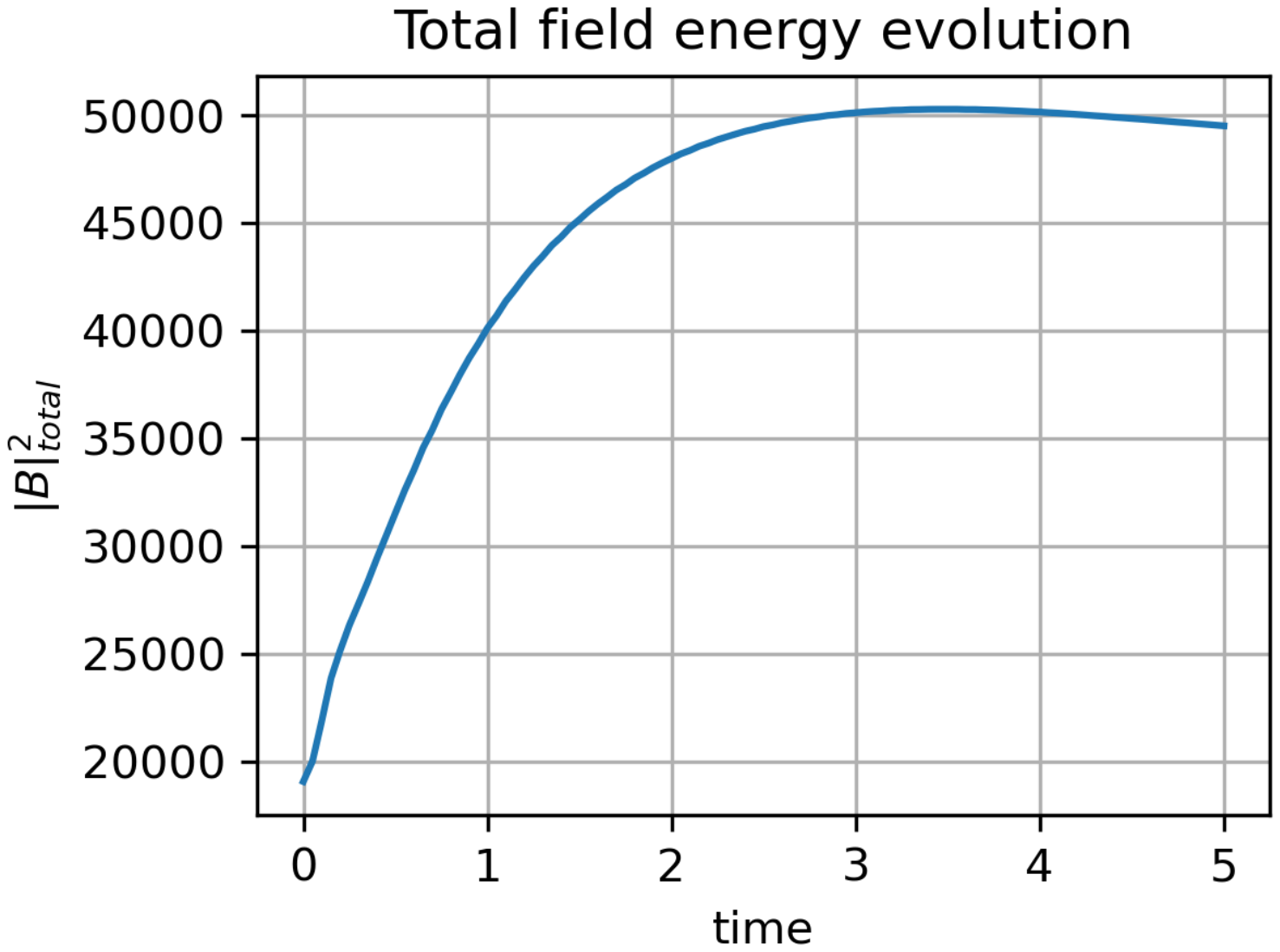}
	\caption{Evolution of the magnetic field energy in the cosmic string wake for a value $\alpha_1 = 0.01$. The magnetic field energy is given in scaled units and hence is dimensionless}
	\label{fig2}
\end{figure}

To understand this, we looked at two different kinds of magnetic field perturbations. We start with an exponentially decreasing magnetic field perturbation. The strength of the magnetic field is proportional to the density of the magnetic field lines. As mentioned before, the cosmic string is moving in the $x$-direction.  Due to the velocity perturbation, the fluid velocity has a component in the $y$-direction, and therefore, the magnetic field lines feel a force along the $y$-direction. Therefore, we start with a sheared perturbation in the direction of motion of the cosmic string. 
Sheared magnetic fields are usually generated by small scale density inhomogeneities and the lengthscale of the magnetic fields are therefore determined by the lengthscale of the inhomogeneities in the cosmic string wake. 
 Fig.\ref{fig2} represents the graph of the total magnetic field with time for a perturbation of $B_x = B_0 e^{-\alpha_1 |y|}$. We call this a sheared magnetic field as the $x$ component of the magnetic field ($B_x$) at any given point has a variation in the $y$-direction. The lengthscale of the variation is given by the parameter $\alpha_1$. We have used $B_0 = 1$.    
\begin{figure}[h]
	\centering
	\includegraphics[width=\linewidth]{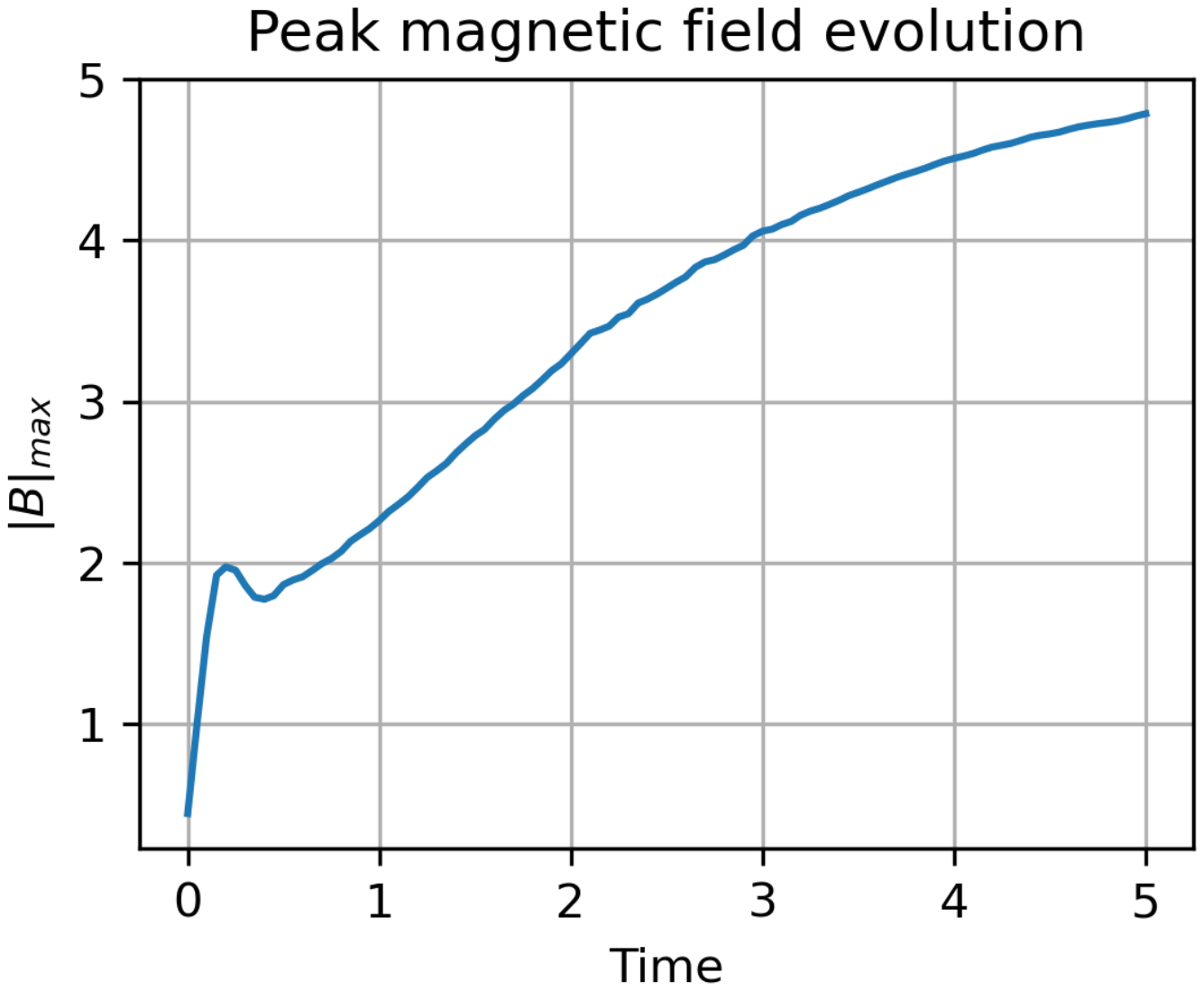}
	\caption{Evolution of the peak value of the magnetic field in the cosmic string wake. The value after a few time steps is five times the initial value of the magnetic field. The magnetic field is given in scaled units and hence is dimensionless. }
	\label{fig3}
\end{figure}

Fig.\ref{fig3} represents the graph of the peak magnetic field evolution with time for $\alpha_1 = 0.01$. As the graphs shows there is an increase in the peak magnetic field value which will lead to the total magnetic energy being enhanced. The figure shows that the peak magnetic field value becomes at least five times the initial value. The peak magnetic value shown on the $y$- axis is the maximum value of the magnetic field at a specific point on the grid.  Similarly, the total energy also becomes five times the initial energy value, the $|B|^2$ on the y - axis is the summation of the magnetic energy density over all the grid points. Together, they indicate that the magnetic field value not only increases at a particular point, it increases over the whole of the cosmic string wake.  So, we see that the net magnetic field gets amplified.

The amplification is a result of the conical nature of the cosmic string metric. There is a velocity perturbation on the particles along the $y$-direction, so the particles move towards the conical space behind the cosmic string. We have given a sheared magnetic field as the magnetic field perturbation. Though we are perturbing the $x$ component, the variation is along the $y$-direction. As the magnetic field lines are being pushed in the $y$-direction by the velocity flow, so our perturbation is enhancing the effect of the compression of the magnetic field lines due to the fluid flow. We find that we get an amplification of the field for all values of $\alpha_1$.

\begin{figure}[h]
	\centering
	\includegraphics[width=\linewidth]{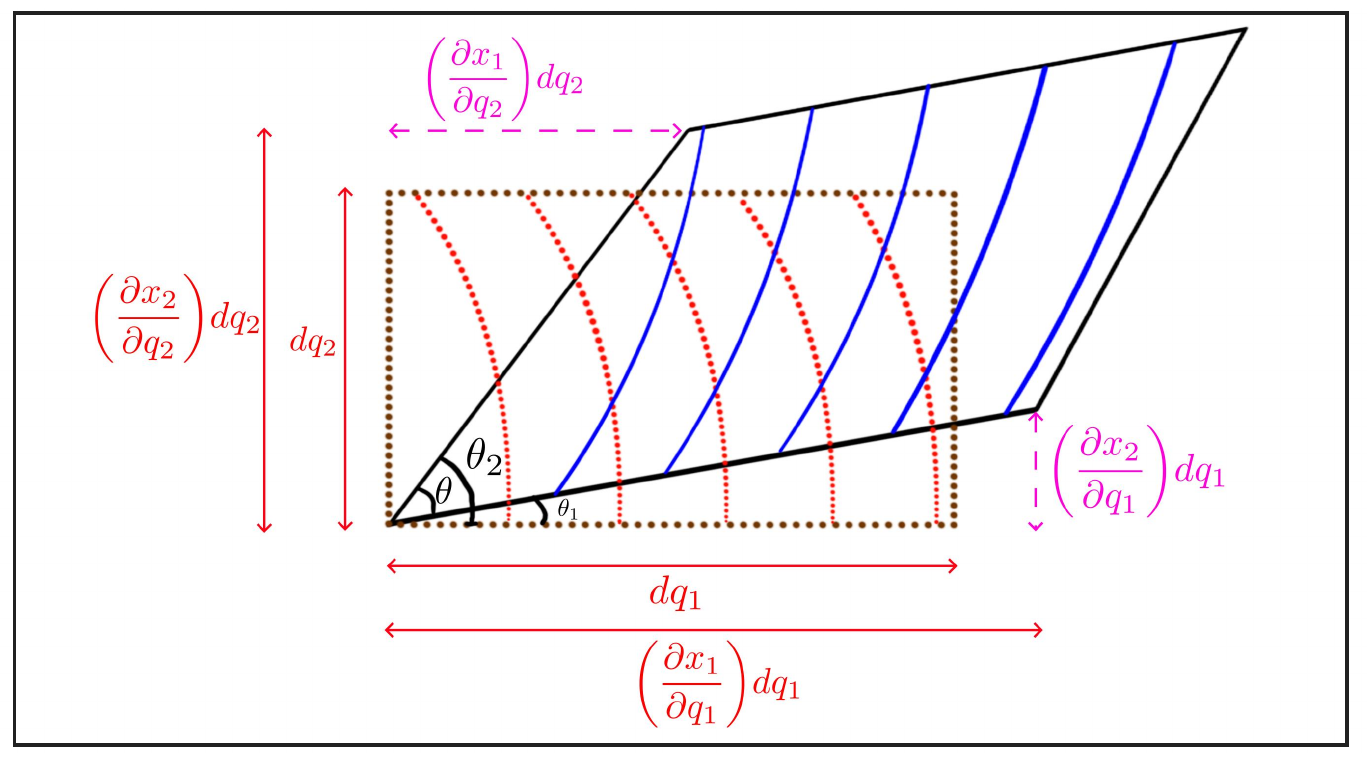}
	\caption{The figure representing the deformation of the magnetic field lines in a fluid flowing around a cosmic string. The point at which the axes $dq_1$ and $dq_2$ meet is the place at which the cosmic string is placed.}
	\label{deformationfig}
\end{figure}

This amplification can be attributed to the deformation tensor of the fluid element. The deformation tensor of the fluid element has been used to explain the amplification of frozen in magnetic fields in ref. \cite{king}. The deformation parameter $D_{ij}$ represents the deformation of an infinitesimal fluid element of initial side length $\delta q_j$, which in this case is deformed due to the deficit angle in the local space of the cosmic string. The deformation tensor is defined by 
\begin{equation}
D_{ij} \equiv \frac{\partial x_i}{\partial q_j}. 
\end{equation}
Fig.\ref{deformationfig} is used to explain how the magnetic field gets amplified for this given magnetic field perturbation. As has already been pointed out in ref. \cite{king}, the magnitude of the magnetic field is affected only when the fluid motion is perpendicular to the magnetic field lines. In this case, for any given value of $x$, the magnetic field varies along $y$ as an exponentially decaying field. This is illustrated in Fig.\ref{deformationfig}. The $q_1$ in the figure is the $x$-axis, while the $q_2$ is the $y$-axis. The angle $\theta$ by which the fluid element is deformed is proportional to the deficit angle of the cosmic string. The comoving volume element is preserved after the deformation, so, $\frac{V^{fin}}{V^{ini}} = D_{ij}$. Since the volume can be related to the density of the fluid, we will have 
\begin{equation}
\frac{V^{fin}}{V^{ini}} = \frac{\rho^{ini}}{\rho^{fin}} = D_{ij}
\end{equation} 
As we are considering frozen in magnetic fields, squeezing or stretching of the field lines will depend on the change in the cross-sectional area. Therefore, we will also have, 
\begin{equation}
\frac{B^{fin}}{B^{ini}} = \frac{A^{ini}}{A^{fin}}
\end{equation}
where, $A$ denotes the initial and final cross-sectional areas through which the magnetic field line passes before and after deformation. If the length perpendicular to this area is referred to as $L$, from volume conservation we will have, 
\begin{equation}
\frac{B^{fin}}{B^{ini}} = \frac{L^{fin}}{L^{ini}}
\end{equation}
The angle between $dq_2$ and $L^{fin}$ is the deficit angle $\theta$. This vector $L^{fin}$ has not been shown on the Fig.\ref{deformationfig} as it is rather difficult to depict in this case. Thus, we will have, 
\begin{equation}
\frac{B^{fin}}{B^{ini}} =  \frac{\rho^{fin}}{\rho^{ini}}\frac{L^{ini}}{L^{fin}}
\end{equation}   
In the case, where $\frac{L^{ini}}{L^{fin}} \approx 1$, the final magnetic field will be determined by the overdensity of the wake.  

In this case, the perturbative field is varying in the $y$-direction. So, we also have off-diagonal elements which are non-zero. The off-diagonal elements correspond to terms like $\frac{\partial x_2}{\partial q_1}$. Here, these terms are non-zero as $\partial x_2$ refers to the change in the $y$-direction corresponding to a small change in the $x$-direction. In all the cases that have been discussed in ref. \cite{king}, the anisotropic collapse has lead to the magnetic field being proportional to the density of the fluid. As the wake has reflection symmetry about the $x$-axis, we will have $D_{12} = D_{21}$, and the other off-diagonal elements can be treated as zero. In such a case, we have 
\begin{equation}
\frac{B^{fin}}{B^{ini}} = \frac{1}{(1 + D_{12})^{1/2}}
\end{equation}   
This will mean that $B \propto \rho^{1/2}$. Similarly in any case, that is considered, the final magnetic field is determined by the final density of the deformed fluid. The deformation tensor we have discussed so far is also referred to as the Zel'dovich deformation tensor \cite{zeldovich}and it depends on the Zel'dovich approximation. Therefore, since in the wake region, we should always have $\rho^{fin} > \rho^{ini}$,  so the magnetic field will always be amplified in the wake of the cosmic string. This will happen as long as the magnetic field remains frozen in the plasma.  We have checked the results for different values of $\alpha_1$ and in all the cases we see an increase in the magnitude of the magnetic field in the wake region.

The magnetic field in the cosmic string wake can have components in both the $x$-and $y$-direction. So, our next case would be to study a perturbation given by $B_y = B_0 e^{- \alpha_2|(x - 800)|}$. The origin is shifted to get a better figure of the final wake. It does not have anything to do with the field magnitude. Here, we depict the lengthscale by the parameter $\alpha_2$. The initial value is still $B_0 = 1$. 
\begin{figure}[h]
	\centering
	\includegraphics[width=\linewidth]{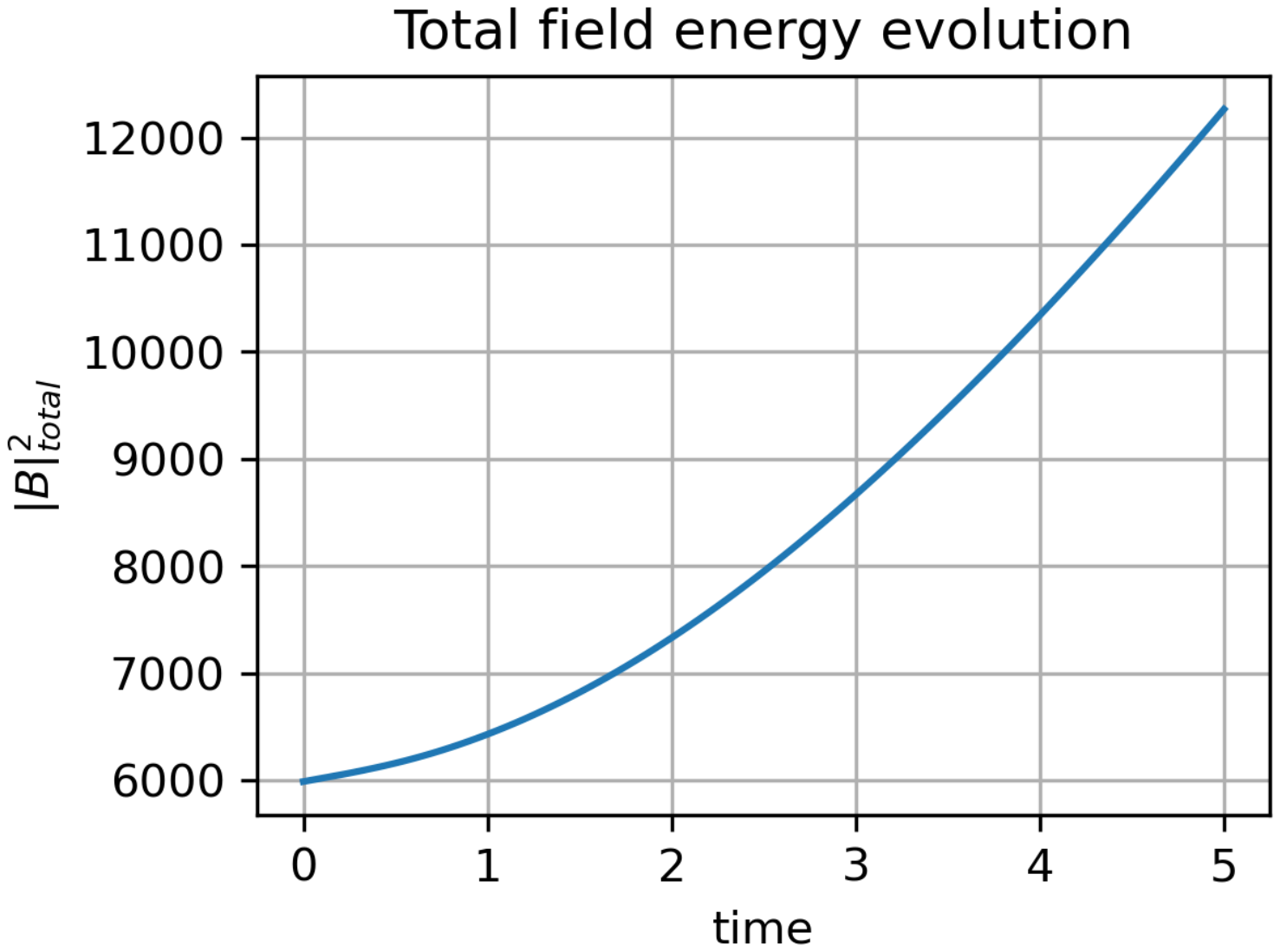}
	\caption{Evolution of the magnetic field energy in the cosmic string wake for a value $\alpha_2 = 0.01$.}
	\label{fig5}
\end{figure}
\begin{figure}[h]
	\centering
	\includegraphics[width=\linewidth]{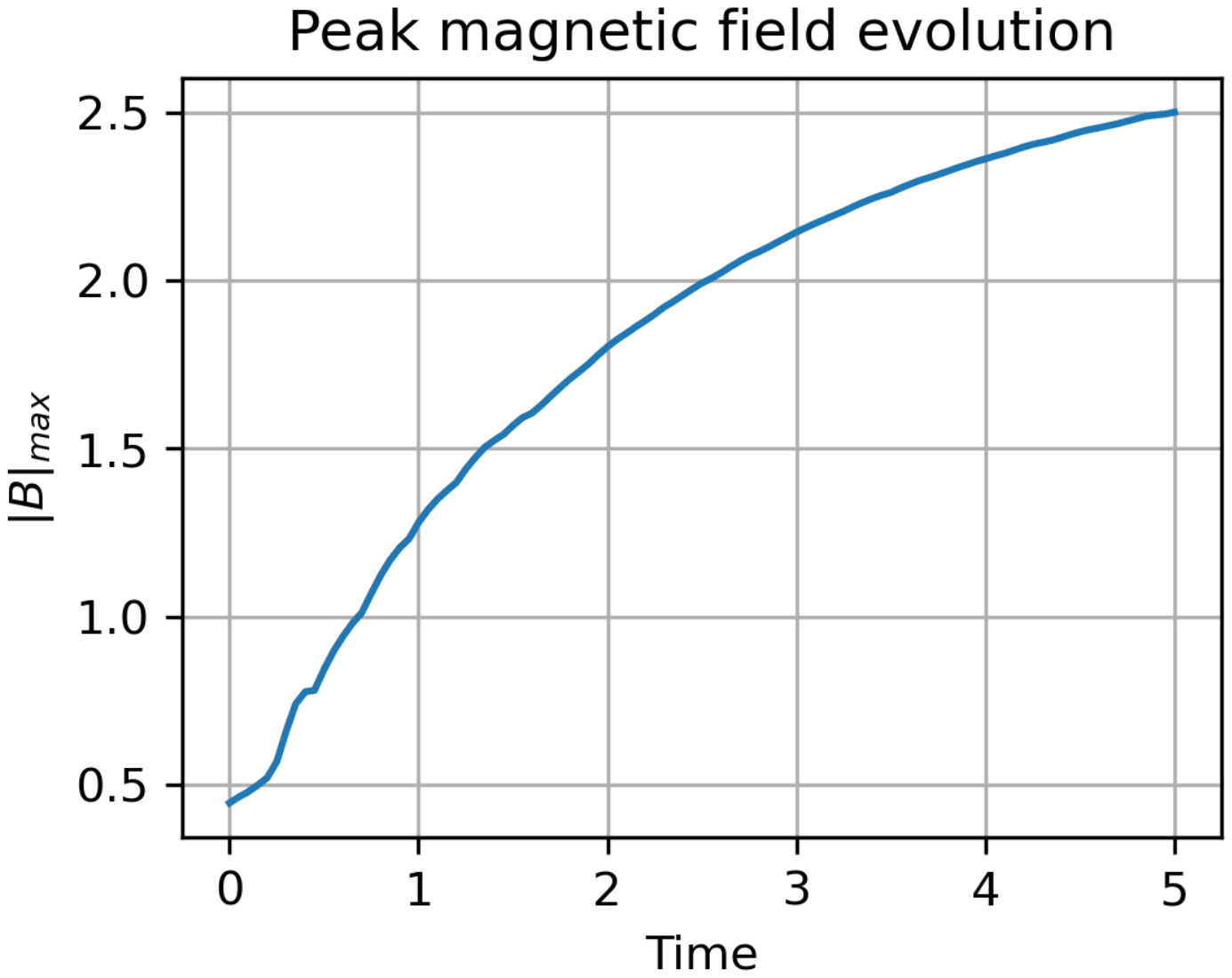}
	\caption{Evolution of the peak value of the magnetic field in the cosmic string wake. The value saturates to about five times the initial value even in this case.}
	\label{fig6}
\end{figure}
As is seen from Fig.\ref{fig5} and Fig.\ref{fig6}, we still get a amplification of the magnetic field. The maximum value is approximately five times the initial value again. In this case too,  the field lines are deformed by the perturbation. As we have taken the flow velocities and the perturbations along the $x$-and $y$-axes, the amplification is similar in both the cases. If the axes were not aligned, we might have got different results for the amplification. In both our cases the off-diagonal elements which are non-zero are driven by symmetry consideration. So, we can show that the magnetic field is proportional to the density of the wake in both the cases. We have repeated the simulation with a different periodic magnetic field and obtained similar results about the magnetic field amplification.  For simplicity, we have taken one of the axes to be aligned with the flow, even if that is not the case, the amplification will still happen, only the symmetry consideration cannot be used to calculate the off-diagonal elements.

\section{Magnetic field decay}
\begin{figure}[h]
	\centering
	\includegraphics[width=\linewidth]{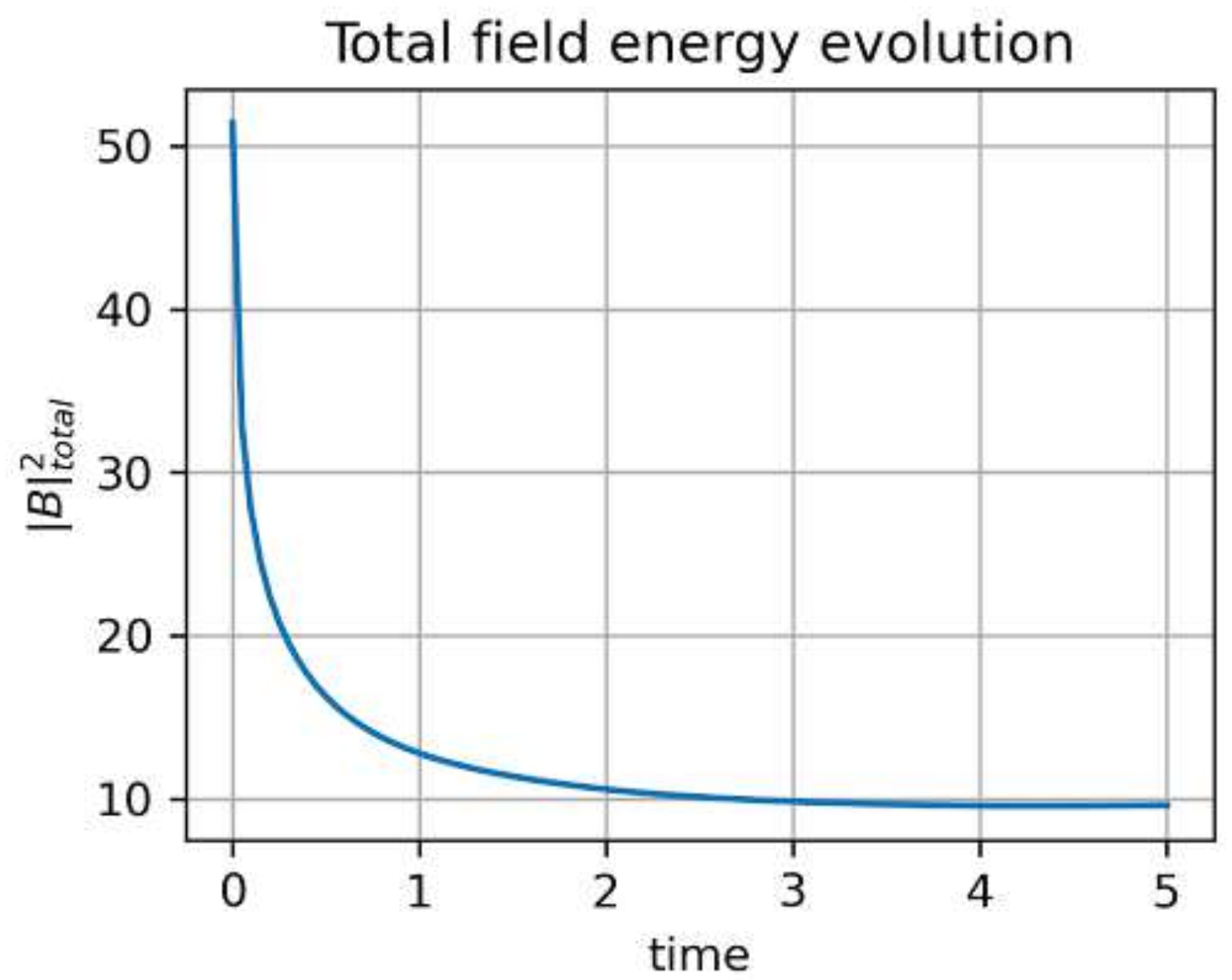}
	\caption{Evolution of the magnetic field energy in the cosmic string wake for a value $\alpha_2 = 1$.}
	\label{fig7}
\end{figure}

\begin{figure}[h]	
	\centering
	\includegraphics[width=\linewidth]{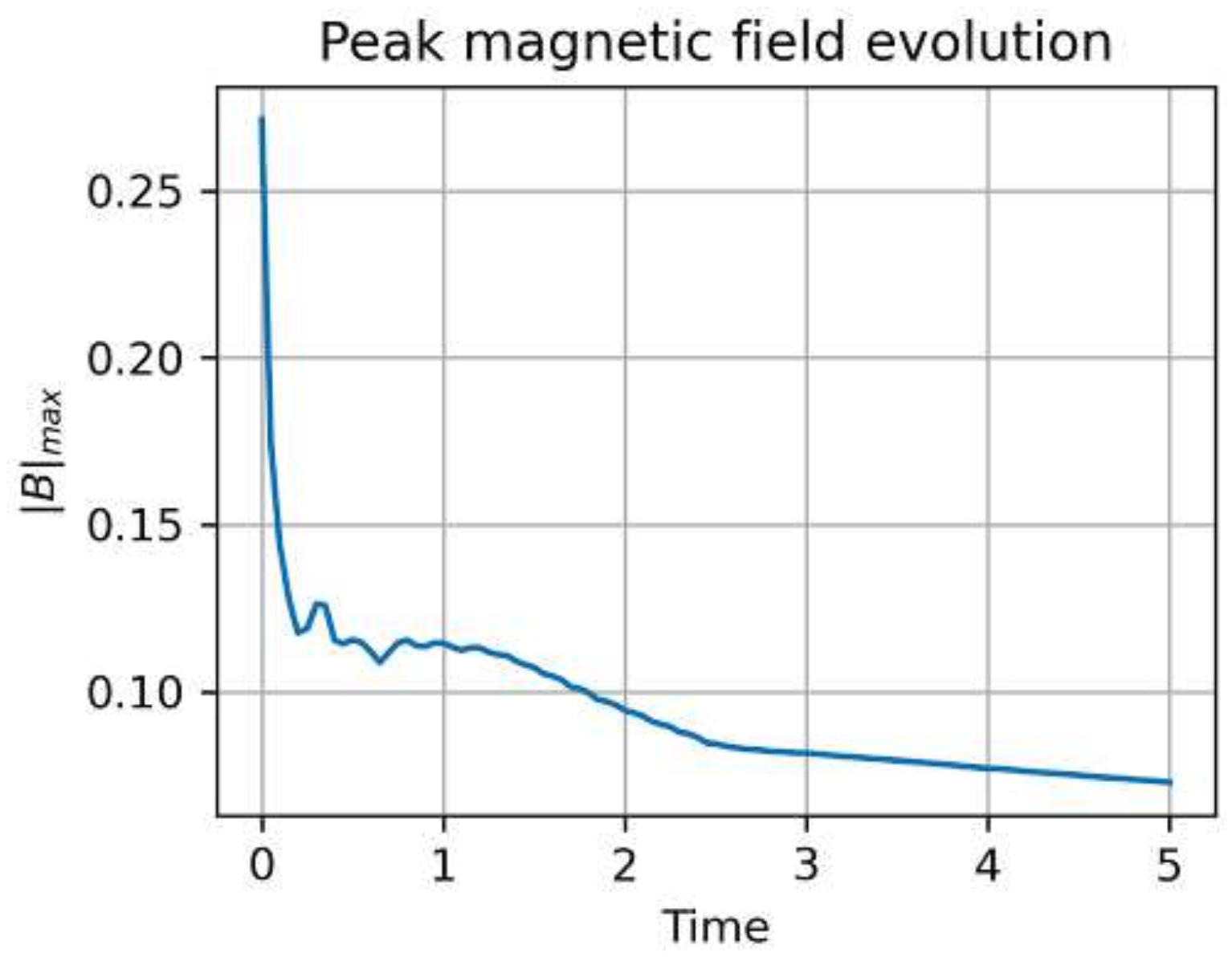}
	\caption{Evolution of the peak value of the magnetic field in the cosmic string wake at a particular point in the wake.}
	\label{fig8}
\end{figure}

While surveying the various values of $\alpha_1$ and $\alpha_2$, we also found that while the field always increases for any value of $\alpha_1$, this is not the case for $\alpha_2$. For $\alpha_2 = 1$, we find that the field decays with time. This is plotted in Fig.\ref{fig7} and Fig.\ref{fig8}. 
So, for a perturbation perpendicular to the flow velocity of the plasma, the lengthscale of the perturbation determines whether the field decays or it gets amplified. As discussed in the previous section, the Alfven's theorem leads to the frozen in magnetic field lines. Due to the deformation of the fluid, the magnetic field lines get squeezed, and the amplification of the field is observed. However, in this case too the fluid is deformed but the magnetic field does not get amplified. This seems to indicate that Alfven's theorem is no longer valid for this lengthscale. Zel'dovich approximation while true for all ideal fluids breaks down under certain conditions. The general assumption on which this is based is that the initial velocity field be irrotational. A magnetic field perturbation perpendicular to the flow will have a rotational component. If this rotational component is not negligible, the approximation will break down.
Alfven's theorem will also become invalid for such a flow.

The breakdown of Alfven's theorem has been studied in connection to ideal plasma flows \cite{eyink}. This breakdown leads to the phenomenon of magnetic reconnection in space plasmas. It has been shown that flux conservation is violated at an arbitrarily small lengthscale ($l_0$) for certain conditions. For such lengthscales, dissipative non-ideal effects will be significant. The question arises how do we identify this lengthscale in our problem. Our numerical simulations indicate that it is related to the value of $\alpha_2$ of the perturbative magnetic field. So, we concentrate on the lengthscale $l \sim 1/\alpha_2$, which we denote by $l_0$. The magnetic field can be considered to be \enquote{coarse-grained} at lengthscale $l_0$ so that at scales smaller than this lengthscale the fluid becomes non-ideal. 

One of the ways to identify this lengthscale analytically, would be to study the particle trajectories of the charged particles in the plasma. For this, we start with obtaining the vector potential of the magnetic field perturbation. Though not unique, the most acceptable vector potential which will give us a magnetic field like $B_y = B_0 e^{- \alpha_2|(x - 800)|}$ is given by $\vec{A} = \hat{k}\frac{B_0}{\alpha_2} e^{- \alpha_2 x} $. We have shifted the origin back to zero for ease of calculation.  Since the vector component does not depend on the coordinates $(y,z)$, their respective momenta should be conserved. Therefore we can take the initial velocities as constant in these directions $v_y (t = 0) = v_0$ and $v_z (t = 0) = u_0$. Based on these constants, we can then obtain the effective potential which determines the trajectories of these particles. The effective potential is given by, 
\begin{equation}
 \Phi(x) = \frac{\omega_B^2}{2 \alpha_2^2} \left( e^{-2\alpha_2 x} - 1\right) + \frac{u_0 \omega_B}{\alpha_2} \left(1 - e^{-\alpha_2 x} \right)
\end{equation}  
Here $\omega_B$ is the gyrofrequency of the charged particle in the given magnetic field. 
This effective potential has an extremum point which will tell us whether the particle motion is bounded and stable or unbounded and unstable. If the particle motion is stable then the Alfven's theorem would hold as the particle paths will be in well defined orbits and the mapping between the initial to the later coordinates remain single valued and continuous. For this the Zel'dovich's approximation holds and the deformation tensor can be used to obtain the amplification of the field. When the particle path becomes unstable, Zel'dovich's approximation no longer holds and the mapping may not be single valued. We will therefore obtain the condition for the extremum to be unstable to obtain the lengthscale at which the Alfven's theorem breaks down.  We find that the extremum is a minima given by, 
\begin{equation}
x_{min} = -\frac{1}{\alpha_2} log \left(-\frac{u_0 \alpha_2}{\omega_B} \right)
\end{equation}
We assume that $\alpha_2$ being small, we can expand the logarithmic term in a series expansion. Considering only the first term of the series we find that $x_{min} \propto\frac{u_0}{\omega_B} $.

So the minimum lengthscale for stability is proportional to the gyroradius as $\omega_B$ is the gyrofrequency of the charged particle. The gyroradius will be given by $r_g = \frac{m  u_0}{qB}  = \frac{u_0}{\omega_B} $.  
Assuming the majority of the particles to be electrons, one can then calculate the gyroradius of the plasma electrons. We find that the gyroradius gives us a good estimate of whether $l_0$ has been reached. Based on the fluid velocity, we calculate an approximate value of the electron gyroradius $r_g$. We find that the magnetic field gets amplified only when the particle gyroradius $r_g$ is smaller than the scale at which the magnetic field perturbation varies, i.e. for $r_g < 1/\alpha_2$. Since in our numerical simulations the approximate value of the gyroradius is of the order of $r_g \sim 1.31$, hence we find that when the value of $\alpha_2 \sim 1$, the field decays instead of getting amplified. This means that the particle motion is no longer stable and it does not have a well defined orbit anymore. Both Zel'dovich's approximation as well as Alfven's theorem break down as the mapping between the initial point in the path to the final point in the path of the particle no longer remains unique and single valued.

Another explanation for this lengthscale can be obtained from the description of the motion of charged particles in terms of the gradient and curvature drift. This approach has been studied in specific sheared force-free magnetic fields \cite{vekstein} before. In this approach, the motion of the electrically charged particles in a perturbed magnetic field has  been studied using the approximate description called the guiding center or drift approximation \cite{northrop}. The charged particle motion in the plasma is decomposed into a nearly circular fast gyromotion about a guiding center and the trajectory of the guiding center itself. Neither of the trajectories are constant for the non-uniform magnetic field. The overall motion is obtained by solving the second-order differential equations for the guiding center. Usually, for better understanding, the equations are separated into parallel and perpendicular components of the given magnetic field. In our present case, the perpendicular direction is the direction along which the field is non-uniform. The velocity in the perpendicular direction  (i.e. along the non-uniform direction) is then given by a gradient drift ($v_g$) and a curvature drift ($v_c$). These are defined by, 
\begin{equation}
\boldsymbol{v_g} = \frac{m v_y^2 c}{2 q B} \left(\mathbf{h} \times \frac{\nabla \mathbf{B}}{B} \right)
\end{equation}  
\begin{equation}
\boldsymbol{v_c} = \frac{m v_x^2 c}{q B R^2} \left(\mathbf{R} \times \mathbf{h} \right)
\end{equation}
Here, $\mathbf{h} = \frac{\mathbf{B}}{B}$ is the unit vector along the magnetic field line and $\mathbf{R}$ is its radius of curvature. From the definition of the gyroradius and the fact that the term $\frac{\nabla \mathbf{B}}{B}$ will be inversely proportional to the lengthscale of the varying magnetic field, we can estimate the order of magnitude of the drift velocity of the particles will be given by $v \times \frac{r_g}{L}$. Here, $L$ is the spatial scale of the non-uniformity of the magnetic field. 
For $r_g << L$, the drift velocity approximation is valid. Since this approximation indicates that the velocity of the magnetic lines can be looked as the velocity of the trajectory of the center of the gyro motion, Alfven's theorem will be valid in this approximation. For $r_g \sim L $, the lengthscale of the variation of the magnetic field will be the same as the gyroradius, so the gyroradius cannot be defined at all for these electrons. This is because the underlying assumption in defining the gyroradius is that the field should be uniform. So, the concept of the guiding center cannot be used to understand the movement of the magnetic field lines. So, for $r_g \geq L $, the magnetic field lines no longer behave as if they are frozen in the plasma.

Thus, we conclude that sheared magnetic fields with lengthscales greater than the gyroradius will be amplified in the cosmic string wake, whereas magnetic fields with smaller lengthscales will decay. So, for sheared perturbations of small lengthscales, Alfven's theorem is no longer valid. This is similar to the situation in the Sweet-Parker model of magnetic reconnection. In that case, two opposite fields close to each other generate a neutral region where the magnetic lines of force can break and reconnect. This is the resistive diffusive region; outside this region, Alfven's theorem of frozen in magnetic field lines still holds while inside this region, the magnetic lines of force break and form an X-type neutral point \cite{reconnection}. Since it has been demonstrated that magnetic reconnection can happen in cosmic string wakes \cite{dilip}, it is important to understand the small-scale dynamics of the magnetic field in the cosmic string wakes. From this study, it appears that the lengthscale of the reconnection region for cosmic string wakes should be less than the gyro radius corresponding to the magnetic field in the wake.

\section{Summary and Conclusions}

In this work, we have studied the evolution of a magnetic field in a cosmic string wake. This is an important thing to understand as in recent times there is a renewed interest in identifying specific signatures of these strings through the generation of Gravitational Waves (GW)\cite{ligo}. While it is possible that GW's may lead to a signal for the cosmic string it is important to look for other signatures too. 
This is especially important as it has been recently established that cosmic string wakes may have magnetic fields within their wakes. Moreover since the width of the wake is quite narrow there is the possibility of magnetic reconnection. This means that apart from GW's, synchrotron radiation as well as GRB's may be additional features to look for when searching for these elusive objects. In view of this it is important to look at the lengthscales associated with the magnetic field evolution in the cosmic string wake.    

In this work, we have found that the magnetic field in the wake region is amplified under certain circumstances of the flow. This is expected as the field lines are frozen in and so the conservation of magnetic flux density usually enhances the field when the fluid flow squeezes the flux lines together. The magnetic field is amplified due to the squeezing of the particle trajectories causing the overdensity in the wake region. Assuming the wake is symmetric about the flow axis, we show that the deformation tensor will have non-zero off-diagonal terms. This leads to the fact that the magnetic field strength will become proportional to the square root of the fluid density. Depending on the orientation of the deformation, the magnetic field can also be proportional to the density. Since the overdensity in the wake is always greater than the initial density of the flow, the magnetic field flux gets amplified in all the different cases.  In this work, we use an exponentially decreasing magnetic field to illustrate our findings.

Our study also shows that if the sheared magnetic field perturbation in the cosmic string wake is aligned to the flow of the plasma past the string, then the magnetic field is always amplified for whatever perturbation is given. It is only when the magnetic field perturbation is perpendicular to the direction of the flow, the decay lengthscale of the perturbed magnetic field determines whether the field gets amplified or whether it decays. We study the lengthscales involved in these cases in detail. We find that the reason of the decay is the breakdown of the Zel'dovich approximation at lengthscale close to the gyroradius. We establish that this happens due to the fact that the effective potential generated by the moving electrons leads to an unstable orbit for the particle. Due to the unstable nature of the motion, the electron ceases to stay in a stable, bound orbit and instead moves away in random directions. In such a case, the Zel'dovich approximation fails as the initial and final 
points of the particles paths are no longer well defined.

Thus we conclude that for a sheared perturbation which is perpendicular to the direction of flow of the fluid, the magnetic field lengthscale determines whether the magnetic field will amplify or decay. We have determined that the length scale below which the magnetic field decays is determined by the gyroradius of the charged particles being considered. If the lengthscale of the magnetic field is much larger than the gyroradius, Alfven's theorem remains valid and we get the amplification of the field. However, if the lengthscale of the magnetic field perturbation is of the same order or less than the gyroradius of the electron then the field decays. We have presented results for a cosmic string moving in a high $\beta$ plasma as in the pre-recombination era the value of $\beta$ is high. The wake formation in our numerical study is also more pronounced for the high $\beta$ plasma. For the numerical part, we have used the OpenMHD code, an open source code, which is available from GitHub.

The magnetic field due to the Biermann mechanism is due to the small-scale inhomogeneities generated in the cosmic string wake. Since the density perturbations of a long cosmic string are of varied lengthscales based on how they are generated, it is important to understand the behavior of the magnetic field for perturbations of different lengthscales. There are cosmic strings with small-scale structures on them which can generate small-scale non-linear density perturbations. It is also known that non-Gaussian density perturbations can be generated by long strings \cite{dacunha}. As long as Alfven's theorem holds, the magnetic field configurations and lengthscales will directly relate to these density perturbations. However, we have also found that for magnetic perturbations with lengthscales of the order of the gyro radius or less, Alfven's theorem breaks down. This in turn can lead to the possibility of magnetic reconnection in the cosmic string wake. Since, in recent times, magnetic reconnection has been seen as a viable possibility in the narrow wakes of the cosmic string, this study will lead to a further understanding of the effect of these magnetic reconnection in the generation of GRB's.

Our study has some limitations as it is constrained to two dimensions only. The current study therefore has one dimensional sheared magnetic fields. The actual wake structure is three-dimensional, so it is also possible that the perturbation is a planar sheared field. The evolution of such perturbations may lead to further interesting results. We have also considered ideal MHD, it is quite possible that the reconnection possibility and the breakdown of Alfven's theorem will be different for resistive MHD. In fact, there might be a higher probability of obtaining magnetic reconnection in that case. These and other aspects of magnetic field evolution in cosmic string wakes need to be studied in more details in the future.

\begin{acknowledgements}
We acknowledge funding from the DST-SERB Power Grant no. SPG/2021/002228 of the Government of India. D.B was supported through the Science Academies Summer Research Fellowship program 2023, D.K is supported by the Power Grant no   SPG/2021/002228, and S.N acknowledges financial support from the CSIR fellowship no 09/414(2001)/2019-EMR-1 given by the HRD, Govt. of India. 
\end{acknowledgements}


\begin{thebibliography}{30}

\bibitem{vishniac}
G. Ponce and E. T. Vishniac, The Astrophysical Journal, 332, 57 (1988).

\bibitem{sornborger} 
Andrew T. Sornborger and Robert H. Brandenberger and Bruce Fryxell and Kevin Olson, The Astrophysical Journal, 482, 22-32 (1996).

\bibitem{stebbins} 
A. Stebbins, S. Veeraraghavan, R.H. Brandenberger, J. Silk, N. Turok, Astrophys. J. 322, 1 (1987).

\bibitem{danos}
Danos, Rebecca J., Robert H. Brandenberger, and Gil Holder.  Physical Review D 82.2 : 023513 (2010).

\bibitem{vachaspati1} 
T. Vachaspati, Phys. Rev. D 45, 3487 (1992). 

\bibitem{vachaspati2}
T. Vachaspati and A. Vilenkin, Phys. Rev. Lett. 67, 1057  (1991).


\bibitem{sovan} 
S. Sau, S. Sanyal, Eur. Phys. J. C 80, 152 (2020).


\bibitem{layek1}
B. Layek, S. Sanyal and A. M. Srivastava, Phys. Rev. D.63, 083512, (2001).

\bibitem{layek} 
B. Layek, S. Sanyal, and A. M. Srivastava.  Physical Review D 67, 083508 (2003).

\bibitem{sovan1}
S. Sau, S. Bhattacharya and S. Sanyal, Eur. Phys. J. C 79 (5), 439 (2019).

\bibitem{Planck}M. Torki,H.  Hajizadeh, M. Farhang, A. Vafaei Sadr, and S. M. S. Movahed,  
 Monthly Notices of the Royal Astronomical Society, 509(2), 2169-2179 (2022).
 
\bibitem{sousa} L. Sousa,  Gravity, Cosmology, and Astrophysics: A Journey of Exploration and Discovery with Female Pioneers. Cham: Springer Nature Switzerland,  213-235, 2023.

\bibitem{cyr}B. Cyr, J. Chluba and S. K. Acharya, Physical Review D 109.12  L121301 (2024).

\bibitem{rogozin}
D. A. Rogozin and L.V. Zadorozhna, Astron. Nachr., 334: 1051-1054 (2013).

\bibitem{soumen2} D. Kumar, S. Nayak and S. Sanyal,  International Journal of Modern Physics D
Vol. 34, No. 1 (2025) 2450066

\bibitem{soumen}
S. Nayak, S. Sau, S. Sanyal, Astroparticle Physics, 146,102805, (2023).

\bibitem{dilip} 
D. Kumar and S. Sanyal, ApJ 944 183 (2023).

\bibitem{DilipGRB} D. Kumar, B. Bose and S. Sanyal, arXiv 2503.19345

\bibitem{king} 
E. J. King and P. Coles, Mon.Not. R. Astron. Soc. 365,1288-1294 (2006).

\bibitem{abhisek}
A. Saha and S. Sanyal, JCAP03, 022, (2018).

\bibitem{dahlburg} 
R. B. Dahlburg, R. Keppens and G. Einaudi, Phys. Plasmas 8, 1697–1706 (2001).

\bibitem{beresnyak} 
A. Beresnyak, The Astrophysical Journal 804.2 :121; (2015).

\bibitem{kim}
Eun-Jin Kim and Angela V. Olinto and R. Rosner, Astrophysical Journal 468, (28 – 50) (1996).

\bibitem{wu}
J. H. P.  Wu, P. P. Avelino, E. P. S. Shellard, and B. Allen,  International Journal of Modern Physics D, 11(01), 61-102, (2002).

\bibitem{hartmann} B.Hartmann and P. Sirimachan Journal of High Energy Physics, Volume 2010, article id.110 (2010).

\bibitem{vilenkin} 
A. Vilenkin, Phys.Rev. D23: 852-857 (1981).

\bibitem{deruelle} 
N. Deruelle, B. Linet, Class. Quantum Grav. 5, 55 (1988).

\bibitem{Maarten}M. van de Meent, Phys. Rev. D 87, 025020 (2013).

\bibitem{rezolla}L. Rezolla, Lecture Notes, Virgo School (Cascina, 2004). 

\bibitem{bertschinger} E. Bertschinger, The Astrophysical Journal, 316:489-496,(1987).

\bibitem{mkverma} 
M. K. Verma, Physics Reports 401.5, pp. 229–380 (2004).


\bibitem{openmhd} 
S. Zenitani, T. Miyoshi, Phys. Plasmas 18, 022105 (2011);S. Zenitani, Physics of Plasmas, 22, Issue 3, 032114 (2015).

\bibitem{gangui} 
A. Gangui, L. Pogosian, and S. Winitzki, Phys. Rev. D 64, 043001  (2001).

\bibitem{zeldovich}Ya. B. Zel’dovich Ya. B.,  Astronomy and Astrophysics, 5, 84 (1970).

\bibitem{eyink} 
G. L. Eyink and H. Aluie, Physica D 223, 82 - 92, (2006). 

\bibitem{vekstein}
G.E. Vekstein, N. A. Bobrova and S. B. Bulanov, J. Plasma Physics, 67, Pg 215 (2002). 

\bibitem{northrop} 
T.G. Northrop and J. A. Rome Phys. Fluids 21, 384 (1978).

\bibitem{reconnection}
Bhimsen K. Shivamoggi, Physics Reports 127,99-184 (1985).

\bibitem{ligo}Pierre Auclair, Konstantin Leyde and Danièle A. Steer, Journal of Cosmology and Astroparticle Physics (04),005 (2023).

\bibitem{dacunha} 
D. C. Neves da Cunha, J. Harnois-Deraps, R. Brandenberger, A. Amara, and A. Refregier, Physical Review D 98, 083015 (2018).





\end{thebibliography}
\end{document}